\documentclass[a4paper,10pt] {article}

\usepackage{amsfonts,amsmath,amssymb,amsthm}
\usepackage[final]{graphicx}
\usepackage{fancyhdr,sectsty}
\usepackage{hyperref,microtype,makeidx}
\usepackage[nottoc]{tocbibind}

\usepackage{textcomp,float}

\usepackage[all]{xy}

\usepackage{color,pstricks}

\usepackage{fullpage}

\usepackage{setspace}
\theoremstyle{plain}

\numberwithin{equation}{section}

\newcommand{\G}{\mathbb G}

\newcommand{\A}{{\vec{A}}}

\title{A Gauge Field Model  of Modal Completion}
\author{Giovanna Citti\footnote{ G. Citti, Dipartimento di Matematica,  Universit\`{a} di Bologna. Bologna, Italy, giovanna.citti@unibo.it,},     Alessandro Sarti\footnote{A. Sarti, 
		 Centre d'Analyse et de Math\'ematique Sociales - EHESS,   Paris, France.  alessandro.sarti@ehess.fr}}

\date{}

\begin{document}

\maketitle
  \begin{abstract}
Perceptual completion of figures is a basic process revealing the deep architecture of low level vision. In this paper a complete gauge field Lagrangian is proposed allowing to couple the retinex equation with neurogeometrical models and to solve the problem of modal completion, i.e. the pop up of the Kanizsa triangle. Euler-Lagrange equations are derived by variational calculus and numerically solved.  Plausible neurophysiological implementations of the particle and field equations are discussed and a model of the interaction between LGN and visual cortex is proposed.

%Scope of this paper is to  show that it is possible to integrate the retinex model of contrast invariance  and the neurogeometrical models of contour completion in order to propose a new model of  completion, which better
%decribes feedback between retina and cortex.
%The interaction between these families of cells is expressed with the classical instruments of gauge field theory, and
%joint  perception of illusory boundaries and pop up of illusory surfaces in the famous Kanitza triangle is explained.
\end{abstract}

\section{Introduction}

Perceptual completion is a low level visual process studied for more than a century, starting  from the pioneers of the phenomenology of the Gestalt \cite{MWert}. The psychologist Gaetano Kanizsa introduced in \cite{Kanizsa1979} a number of stunning examples of images allowing to clearly perceive the phenomenon of pop up of illusory figures. For example in fig. 1) a triangle with curved boundaries is perceived out of the three pac-men inducers. Kanizsa called this pop up effect "modal completion" because the illusory figure and its boundaries is really perceived with the modality of vision, while the three pac-men are completed to disks with "a-modal completion", meaning that they remain invisible, partially masked by the triangle.
\begin{figure}[H] \label{kanizsa}
\begin{center}
\includegraphics[width=7.5cm]{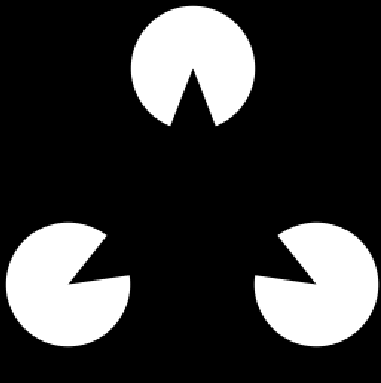}
\end{center}
\caption{The Kanizsa triangle with curved boundaries. Notes the pop up of the illusory triangle out of  the three pacman inducers. }
\end{figure}
%Many mathematical models of perceptual completion have been proposed in the last 3 decades, particularly regarding "amodal completion" and often redefined as inpainting by the authors \cite{BertalmioSapiro} \cite{MorelMasnou}. 

A starting point to afford the task of modal completion is to consider illusory boundaries \cite{PetryMeyer1987, RingachShapley, Kanizsa1979}  and a number of mathematical models have been proposed on this topic. The celebrated model of elastica has been introduced by D.Mumford in \cite {Mumford94} to take into account curvilinear illusory boundaries. Williams and Jacobs proposed a stochastic version of completion fields \cite{WilliamsJacobs}. Recent models of boundary completion are based on the neurogeometrical structure of the visual cortex and they showed a strong explicative power of  perceptual completion. The first neurogeometrical model  has been introduced by Petitot and Y.Tondut  \cite{PetitotTondut} to describe the
functional architecture of the visual cortex with instruments of differential geometry. In particular in  \cite{PetitotTondut} the hypercolumnar structure of the
simple cells organization, responsible for contour detection, is modelled  as a fiber bundle   (see also section 2.2  below). The model has been further developed by Citti and Sarti
\cite{CittiSarti}, \cite{SartiCittiPetitot} who proposed to interpret the whole fiber bundle as the group of position and orientation with a subriemannian metric. This metric allows to reconstruct  rectilinear or curved illusory boundaries.
Other models of  boundary completion have been developed  in the same space \cite{AugustZucker, BenShaharZucker, GuyShahar, RenMalik, Duits1,  Duits2, HladkyPauls, WilliamsJacobs}. 

The problem of modal completion of both boundaries and figures together has been much less covered in literature. 
In \cite{NMS} modal completion has been achieved by non-linear functional minimization by means of combinatorial techniques.
In \cite{Sarti1}\cite{Sarti2} it was proposed a technique to construct the Kanizsa triangle by minimization of an area functional measured with respect to a metric induced by the image. In both models \cite{NMS} and  \cite{Sarti1} a complete boundary/figure reconstruction was provided but a correct filling in of figures with the perceived brighness is still missing.

In this paper we would like to introduce a formal field theory of low level vision, able to afford the
problem of modal completion. The theory will couple two models of low level vision, i.e. a neurogeometrical model of boundary completion and  the celebrated retinex algorithm as a model for filling in.
In particular we will show that it is possible to integrate the retinex  model 
of \cite{Georgiev1} with the neurogeometrical approach of  \cite{CittiSarti}
and to propose a new model of  modal completion, based on complete contrast invariance.
The two equations will act as a particle and a field term of a complete Gauge field theory.

%About the celebrated retinex model, it has been introduced in  \cite{LandCann}\cite{Land} to explain   lightness perception, i.e. the phenomenon causing a  gray patch to appear  brighter  when viewed against a dark background,  and darker  when viewed  against a bright background. Here, we are more interested in his capacity of filling in figures from boundaries.
%After its introduction this model has inspired a
%wide range of  improvements and new models have been proposed \cite{BrainardWandell,McCann,LeiZhouLi,Provenzi,Hurlbert, Kimmel}.
%Particularly  Georgiev \cite{Georgiev1, Georgiev2} applied the fiber bundle framework used in the neurogeometrical setting to the retinex model, capturing the fact that image processing is invariant with respect to change of contrast, and then it is expressed by covariant derivatives.

The paper is organized as follows. In section 2 we recall the main properties of the Retinex algorithm and the neurogeometrical model, and reinterpret them with instruments of gauge field theory. In section 3 we couple the models by introducing a complete gauge field Lagrangian. The corresponding Euler Lagrange equations are calculated by variational calculus. In section 4 the Euler Lagrange equations are solved, providing results on the pop up of the Kanizsa figure. In Section 5 a plausible neural implementation of the model is proposed and discussed. 

\section{The retinex algorithm and the neurogeometrical model}

In this section we recall the main properties of two previously recalled models retinex and the neurogeometrical one. The retinex algorithm has been inspired by the functionality of the retina in detecting image gradients and implementing contrast invariance. The second one has been inspired by the ability of the cortex to detect and complete  boundaries.  We will provide here a short description of the two processes, stressing the similarity of the mathematical instruments adopted by both.
\subsection{A mathematical interpretation of the Retinex algorithm}
The celebrated retinex model has been introduced in  \cite{LandCann}\cite{Land} to explain   lightness perception, i.e. the phenomenon causing a  gray patch to appear  brighter  when viewed against a dark background,  and darker  when viewed  against a bright background. Here, we are more interested in his capacity of filling in figures from boundaries.
After its introduction this model has inspired a
wide range of  improvements and new models have been proposed \cite{BrainardWandell,McCann,LeiZhouLi,Provenzi,Hurlbert, Kimmel}.

In Horn's work \cite{Horn}  the authors proposed a physically based algorithm, which recovers the
reflectance $f$ of an image $I$ as
\begin{equation}\label{logretinexclassical}
 \Delta log f = \Delta log I.
\end{equation}

In \cite{Morel}, it has been proved that the original Retinex algorithm can be equivalently espressed by the same equation.
Precisely then Retinex is equivalent to a
Neumann problem for a linear  equation. The  equation is identical
to the Poisson equation for image editing   proposed in Perez
et al. \cite{Perez}.
 In  \cite{Georgiev1} a new interpretation was given in terms of covariant derivatives and fiber bundles. Indeed
setting \begin{equation}\label{AgradI}\A= \nabla I/I\end{equation}
equation \ref{logretinexclassical} can be considered the Euler Lagrange equation of the functional
\begin{equation}\label{ffunctional}
\tilde{F}=\int \frac{|\nabla f -\A f |^2}{f^2} dx dy.
\end{equation}
This functional is invariant with respect to the transformation
\begin{equation}
f \to fI, \quad \A \to \A + \frac{\nabla I}{I}
\end{equation}
so that the choice $ \A = \frac{\nabla I}{I}$ is compatible with the transformations which
leaves invariant the functional. The quantity
 $\nabla f - \A f$
can be interpreted as a covariant derivative.

Also note that, setting
\begin{equation}\label{logset}
 \phi= log f , \quad h= log I,
\end{equation}
equation \ref{logretinexclassical} further simplify as
\begin{equation}\label{logfieldset}
 \Delta\phi =  \Delta h
\end{equation}
and with the same choice as before: $\A = \nabla I /I=\nabla g$ the functional becomes
\begin{equation}\label{functional}
F=\int |\nabla \phi - \A |^2 dx dy=\int |\nabla \phi - \nabla h |^2 dx dy,
\end{equation}
while the transformations which leave invariant the operator become
\begin{equation}
\phi \to \phi + h, \quad \A \to \A + \nabla h.
\end{equation}

\subsection{A neurogeometrical  model for boundary completion}
Let us recall here the neurogeometrical model of boundary completion proposed by  \cite{CittiSarti}.
 The model mimic the ability of simple cells of detecting boundaries 
and level lines of images, and to complete missing boundaries. 

The retina can be modelled as a 2D plane, whose points will be denoted by $(x,y)$. 
Over each retinic point $(x,y)$ the primary visual cortex implements a whole fiber of cells, each one sensible to a 
specific direction $\theta$. Hence the set of simple cells is identified with the 3D space $R^2 \times S^1$. 
A visual stimulus will be expressed as an image $I(x,y)$ defined on the retinal 2D plane, and we will denote $\bar \theta(x,y)$  the orientation  of his level lines   at every point. 
In presence of a visual stimulus, at every point $(x,y)$ the simple cell sensible to the 
orientation $\bar \theta(x,y)$ will be maximally activated. Hence  the set of 
activated cells defines a surface  in the 3D cortical space  $R^2\times S^1$
$$\Sigma=\{(x,y,\theta): \theta = \bar \theta (x,y), |\nabla I (x,y)|>C \} .$$ The condition on the gradient of $I$ is a treshold, which ensures that 
the function $\bar \theta$ is well defined around boudaries of the image. If we set $\bar H(x,y,\theta)= \theta - \bar \theta (x,y)$, $\Sigma$ will be identified with the 0-level set of $\bar H$: $$\Sigma=\{ \bar H(x,y,\theta) = 0,  |\nabla I (x,y)|>C\}  .$$ 
Simple cells are connected one to the other by the so called cortico-cortical connectivity. 
 This connectivity is strongly anysotropic, and a cell  located at a point $(x,y)$ and sensible to an orientation $\theta$ 
mainly propagate in the direction of its orientation $\theta$. 
More precisely the connectivity allows a propagation of the signal in the  $R^2\times S^1$ along the integral curves of the vector fields 
\begin{equation}\label{fields} X_1 = cos(\theta) \partial_x + \sin(\theta) \partial_y,\quad  X_2 = \partial_\theta. \end{equation}

Propagation along the cortical connectivity 
seems to be at the basis of the process of boundary completion. Indeed the lifted surface $\Sigma$  
is not defined on the whole space, but only over the region where boundaries or level lines are detected. 
The joint action of orientation detection and cortical propagation along the vector fields 
completes the surface extending it on the set $\{|\nabla I|< C\}$. In \cite{CittiSarti} it is shown that it can be  expressed as the solution of the 
minimal surface equation 
\begin{equation}\label{minimal} X_1 \Big( \frac{X_1 H}{\sqrt{|X_1 H|^2 + |X_2 H|^2}}  \Big) + X_2 \Big(\frac{X_2 H }{\sqrt{|X_1 H|^2 + |X_2  H|^2}}\Big) =0, \text{ on }| \nabla I|<C\end{equation}
with internal boundary condition 
$$H=\bar H \text{ on }|\nabla I| =C.$$ This last condition ensures that the existing boundaries are preserved, 
while the orientations of illusory boundaries or level lines are recovered as the 0 level set of the solution $H$.

%Together with boundary reconstruction, the model  \cite{CittiSarti}  performs the reconstruction of the figure, 
%via a Laplace Beltrami operator on the surface. 
This model  performs completion of boundaries, giving rise to illusory contours, and of level lines, giving rise to amodal completion, as in the case of the macula cieca. But it is unable to perform filling  
when the level lines of the image are parallel to the missing or occluded regions, as in the case of modal completion of the Kanizsa triangle. 

%(see for example \cite{HP} for a detailed proof). 

We explicitly note that this is a model of cortical 3D space of position and orientations. 
On the other hand the model defines a surface, that can be also expressed as a graph on the 2D space. 

In facts, if we  project the previous two vector fields on the $x,y$ plane, we end up with a unique derivative
\begin{equation}\label{field1} X_{1 \theta} = cos(\theta (x,y)) \partial_x + \sin(\theta(x,y) ) \partial_y = <\nabla, (\cos(\theta (x,y)), \sin(\theta (x,y)))> \end{equation}
since the projection of the vector $X_2$ on the same plane is $0$. 
We explicitly note that the vector $X_{1 \theta}$ here is only formally similar to the vector  $X_{1}$  in (\ref{fields}). Indeed  $\theta(x,y)$  in \ref{field1} is a function 
while in  (\ref{fields}) $\theta$ was simply  an axis of the 3D space.

The minimal surfaces equation can now be represented as the equation for a graph of $\theta(x,y)$, which is a function of the two variables $(x,y)$ alone. Hence the equation becomes:
\begin{equation}\label{minimal1} X_{1 \theta} \Big( \frac{X_{1 \theta} \theta(x,y) }{\sqrt{|X_{1 \theta} \theta(x,y) |^2 +1}}  \Big)  =0, \end{equation}
where $\theta$ coincides with $\bar \theta$ on the existing boundaries. 
Taking explicitly the derivative, the equation is equivalent to 
 \begin{equation}\label{minimal1} X_{1 \theta}^2 \theta(x,y) =0, \end{equation}
This equation can be interpreted as a second order directional derivative, in the direction $(\cos(\theta), \sin(\theta))$. 
Hence the norm, of the gradient coincides with the directional derivative:
\begin{equation}\label{norm}|v|_\theta^2= \Big( <v, (\cos(\theta), \sin(\theta))> \Big)^2 =\cos^2 (\theta) v^2_x + \sin^2 (\theta) v_y^2 + 2 \cos(\theta) \sin(\theta)v_x v_y.\end{equation}
The projected norm of the gradient of $\theta$ reads:
$$ |\nabla \theta |_\theta^2= |\cos(\theta) \partial_x \theta +\sin(\theta) \partial_y \theta|^2  = $$$$=\cos(\theta)^2|\partial_x \theta|^2 + 2 \cos(\theta)\sin(\theta) \partial_x \theta \partial_y \theta + \sin(\theta)^2 |\partial_y \theta|^2. $$
It's easy to check that the second order equation \eqref{minimal1}  is simply the Euler Langrangian equation of the Dirichlet  functional \begin{equation}\label{tetafunctional}\int |\nabla \theta(x,y) |_\theta^2dx dy. \end{equation}
Hence minima of this functional 
give rise to the same minimal graphs proposed in \cite{CittiSarti} for boundary propagation (see also  \cite{Barbieri} for a detailed proof).

%Let us explicitly note that the graph of $\theta$:
%$$\Sigma=\{(x,y,\theta(x,y)\}$$ is a surface in  $R^2\times S^1$  invariant with respect to rotation and translation of the whole 3D space.
%In addition, the vector fields have been defined as the 2D projection of left invariant vector fields.
%If we fix a point $(x_0, y_0, \theta_0)$  and denote respectively $ R_{\theta_0}$ $T_{x_0, y_0} $ rotations and translations, then
%this amounts to say that the transformation which leaves invariant the derivative $X_{\theta} \theta $  is
%$$(x,y) \rightarrow  R_{\theta_0}T_{x_0, y_0} (x,y) \quad \theta(x,y) \rightarrow \theta\Big( R_{\theta_0}T_{x_0, y_0} (x,y)\Big) + \theta_0.$$
%However note that the model is invariant only with respect to constant coefficients transformations.

\section{The Gauge field model}

\subsection{The Lagrangian}

We will provide a description of the low level vision process taking into account both the retinex model and the cortical neurogeometry. The task will be accomplished by considering the retinex model of Section 2.1 as the particle term and the cortical model of Section 2.2 as the field term of a classical gauge field  theory.

In this way we will obtain an analogous of the classical theory of electromagnetism where both the particle and the fields are the unknown of the problem.
Indeed, instead of equation (\ref{functional}) we propose a complete Lagrangian, sum of three terms:
a particle term, an interaction term and a field term.

The particle term is
\begin{equation}
\mathcal{L}_1 =  \int |d \phi - d h |^2 dx dy
\end{equation}
and  is directly inspired by the retinex model \ref{functional}: it describes the reconstruction of the image from image  boundaries. As described above it implements the perceptual invariance with respects to contrast.
The next term describes the interaction beetween particle and field and it is again a retinex term acting not anymore on existing boundaries but on existing and illusory boundaries marked by the gauge field $\A$, that now is unknown:

\begin{equation}
\mathcal{L}_2 =  \int |d \phi - \A |^2 dx dy.
\end{equation}
It expresses the reconstruction of the image from the old and new boundaries  explaining perceptual figure  completion, by  keeping contrast invariance properties. We explicitly note that in the minimization process $\A$ will have the direction of $\nabla \phi$, hence it will tend to be orthogonal to the existing boundaries or level lines. The orthogonality condition can be expressed in terms of directional derivatives in the direction of $\A$.

%\begin{equation} \label{GF}
%\tilde{{\mathcal{L}}}_3=\int |\nabla \A |^2  dx dy.
%\end{equation}
%In our case it expresses the propagation of existing contours allowing the creation of subjective contours, and it will be modified accordingly to the
%the model presented in Section in 2.2, making use of the previously recalled subriemannian metric.
%In fact, propagation is expected in the direction of the boundary, which is the direction of $\A^\bot$. 
%Since the vector  $\A=(A_x,A_y)$  is not unitary, we will simply
%normalize it to reduce to the norm defined in (\ref{norm}).
%The induced squared norm of $v$ reads:
%$$|v|_\A^2= \Big( \frac{<v, \A>}{\sqrt{A_x^2 +A_y^2}} \Big)^2 =\frac{(A_x v_x  +  A_yv_y)^2 }{A_x^2 + A_y^2}= \frac{A_x^2 v_x^2 + 2 A_xA_y v_xv_y +  A_y^2 v_y^2 }{A_x^2 + A_y^2}.$$
%
%This norm will be choosen to compute $\nabla \A$
%
%$$ \mathcal{L}_3=  \int |\nabla \A|_{ \A}^2    dx dy= \int \frac{\A_x^2 |\partial_x   \A |^2 + 2 A_xA_y \partial_x   \A \partial_y   \A +\A_y^2 |\partial_y \vec \A |^2 }{\A_x^2 + \A^2_y}$$
%The resulting  functional $\mathcal{L}=\mathcal{L}_1+\mathcal{L}_2 + \mathcal{L}_3$ is then
%\begin{equation}\label{functionalnorm}
%\mathcal{L}=\int |\nabla \phi - \nabla g |^2 dx dy+\int |\nabla \phi - \A |^2  dx dy + \int |\nabla \A|_{ \A}^2    dx dy.
%\end{equation}
%
%
%
%\subsection{The invariance property of the functional}
%

Finally the gauge field term will be analogous to the one of classical fields theories:
\begin{equation} \label{GF}
\tilde{{\mathcal{L}}}_3=\int |d \A |^2  dx dy.
\end{equation}
In our case it expresses the propagation of existing contours allowing the creation of subjective contours, and it will be modified accordingly to the
the model presented in Section in 2.2, making use of the previously recalled subriemannian metric.
In fact, propagation is expected in the direction of the boundary, which is orthogonal to $\A$: $\A^\bot=(-A_y, A_x)$.
Since this  vector is not unitary, we will 
normalize it to reduce to the norm defined in (\ref{norm}).
The induced squared norm of $v$ reads:
$$|v|_\A^2= \Big( \frac{<v, \A>}{\sqrt{A_x^2 +A_y^2}} \Big)^2 =\frac{(-A_y v_x  +  A_xv_y)^2 }{A_x^2 + A_y^2}= \frac{A_y^2 v_x^2 - 2 A_xA_y v_xv_y +  A_x^2 v_y^2 }{A_x^2 + A_y^2}.$$

Equivalently, if we call \begin{equation}\label{matrixg}G=(g^{ij})_{i,j = 1,2}=\left(\begin{matrix}
\frac{A_y^2}{A_x^2 + A_y^2} & \frac{-A_xA_y}{A_x^2 + A_y^2}\\ \frac{-A_xA_y}{A_x^2 + A_y^2}&\frac{A_x^2}{A_x^2 + A_y^2}
\end{matrix}\right),\end{equation}

this norm can be computed as 
$$|v|_\A^2 = <Gv, v>,$$
hence the norm is formally the norm associated to the matrix $G=(g^{ij})$. In this setting $G$ is not invertile. On the contrary, in the riemannian setting, $G$ is invertible, and it has the role of the inverse of the metric of the space. When needed we will assume to introduce a small perturbation which makes its determinant non zero:
$$G_\epsilon=\left(\begin{matrix}
\frac{A_y^2 + \epsilon^2 A_x^2}{A_x^2 + A_y^2} &\frac{-A_xA_y(1- \epsilon^2)}{A_x^2 + A_y^2}\\ \frac{-A_xA_y(1- \epsilon^2)}{A_x^2 + A_y^2}&\frac{A_x^2 +\epsilon^2 A_y^2}{A_x^2 + A_y^2}
\end{matrix}\right).$$
We recall that the differential of $\A$ is independent of the norm chosen, and it is the usual curl operator: 
$$d\A = curl (\A) =\partial_x A_y - \partial_y A_x.$$

%This norm will be choosen to compute $\nabla \A$
%
%$$ \mathcal{L}_3=  \int |\nabla \A|_{ \A}^2    dx dy= \int \frac{\A_x^2 |\partial_x   \A |^2 + 2 A_xA_y \partial_x   \A \partial_y   \A +\A_y^2 |\partial_y \vec \A |^2 }{\A_x^2 + \A^2_y}$$
The resulting  functional $\mathcal{L}=\mathcal{L}_1+\mathcal{L}_2 + \mathcal{L}_3$ is then
\begin{equation}\label{functionalnorm}
\mathcal{L}(h, \phi, \A)=\int |d \phi - d h |^2 dx dy+\int |d \phi - \A |^2  dx dy + \int |d\A|_{ \A}^2    dx dy.
\end{equation}

\subsection{The Euler Lagrange equation}

The 
Euler Lagrange Equation of the functional \ref{functionalnorm} becomes:
\begin{equation} \label{Euler}\Bigg\{\begin{matrix}  \Delta \phi =\frac{1}{2}( \Delta h + div (\A)) \\ d_\A^*d\A =- \nabla_\A \phi + \A. \end{matrix} \end{equation}

These equations clearly have the meaning inherited by the corresponding terms of the functional:
the first term is the particle equation that takes the two boundary terms (i.e. the rescaled Laplacian $\Delta g $ of the original image and the contribution $div (\A)$ generated by the gauge field $\A$ and performs a reconstruction of the image by filling in objects. Note that the two terms $\mathcal{L}_1$ and $\mathcal{L}_2$ which generalize the retinex
functional give rise to an unique particle equation.

Note that  $\A=(A_x, A_y)$ is a vector, hence the equation for $\A$  is indeed a system. 
In the same equation the directional gradient is defined as
$$ \nabla_\A \phi = G \nabla \phi,$$

where $G$ is defined in (\ref{matrixg}). 

The term $d_\A^*d$ coincides   with $   \nabla_\A ^\bot \;  curl (\A)$
(see Appendix A) and precisely:
$$d_\A^*d\A=   \nabla_\A ^\bot\;  curl (\A) =- (\nabla_\A (\partial_xA_y- \partial_yA_x))_y  dx + (\nabla_\A (\partial_xA_y- \partial_yA_x))_x  dy.$$
(Here $A_x$ denotes the $x$ component of $\A$, not the derivative). 
In Appendix A we provide its explicit expression as  sum of three terms: 
$$ d_\A^*d\A =  \left\{ \begin{matrix} \Delta_{\A} A_x    - \partial_x a + T_x(\A) \\  \\
 \Delta_{\A} A_y     - \partial_y a + T_y(\A)   \end{matrix}\right.$$
The terms $\Delta _\A$, $T_x(\A)$,  $T_y(\A)$ are defined in (\ref{T12h}). Precisely $\Delta _\A = div (\nabla_\A)$ is the directional Laplacian associated to the considered metric.
 $ T_x(\A),  T_y(\A)$ are advection terms, with coefficients depending on the metric, and 
$$a=  g^{12} \partial_{x} A_y +  g^{22}\partial_{y} A_y   + g^{11} \partial_{x }A_x   + g^{12} \partial_{y }A_x.$$

The field equation on $\A$ propagates the gradient of the image, in the subriemmanian metric, and allows to recover existing and subjective boundaries. We explicitly note that the equation is of second order in the variable $\A$. In general functional of higher order are necessary to obtain completion of 
curved boundaries (as for example in the model of elastica \cite{Mumford94}). However here the field $\A$ has 
the role of approximating the gradient of the image, following the lagrangian $L_2$, hence its second derivatives express third derivatives of the image function.

%$$  \Delta _\A\A =  \Bigg(\begin{matrix}\Delta _\A A_x \\\Delta _\A A_y\end{matrix}\Bigg)=
% \Bigg(\begin{matrix}\frac{A_y^2\partial_{xx} A_x - 2 A_xA_y \partial_{xy} A_x +  A_x^2 \partial_{yy} A_x }{A_x^2 + A_y^2}
%\\
% \frac{A_y^2\partial_{xx} A_y -2 A_xA_y \partial_{xy} A_y +  A_x^2 \partial_{yy} A_y }{A_x^2 + A_y^2 }
%\end{matrix}\Bigg)- 
%  \Bigg(\begin{matrix}
% \partial_x h
%\\
%\partial_yh
%\end{matrix}\Bigg).$$$$+
%  \Bigg(\begin{matrix}
% -\partial_x\Big( \frac{A_xAy }{A_x^2 + A_y^2}\Big)\partial_{x} A_y
% +\partial_x\Big( \frac{A^2_x }{A_x^2 + A_y^2}\Big)\partial_{y} A_y
% +\partial_x\Big( \frac{A^2_y }{A_x^2 + A_y^2}\Big)\partial_{x} A_x
%- \partial_x\Big( \frac{A_xAy }{A_x^2 + A_y^2}\Big)\partial_{y} A_x
%\\
%-\partial_y\Big( \frac{A_xAy }{A_x^2 + A_y^2}\Big)\partial_{x} A_y
% +\partial_y\Big( \frac{A^2_x }{A_x^2 + A_y^2}\Big)\partial_{y} A_y
% +\partial_x\Big( \frac{A^2_y }{A_x^2 + A_y^2}\Big)\partial_{x} A_x
%- \partial_y\Big( \frac{A_xAy }{A_x^2 + A_y^2}\Big)\partial_{y} A_x
%\end{matrix}\Bigg).
%$$
%	
%where $h=-\frac{A_xAy }{A_x^2 + A_y^2}\partial_{x} A_y
% + \frac{A^2_x }{A_x^2 + A_y^2})\partial_{y} A_y
%+ \frac{A^2_y }{A_x^2 + A_y^2}\partial_{x} A_x
%- \frac{A_xAy }{A_x^2 + A_y^2}\partial_{y} A_x$.

\subsubsection{Nonlinearity of the equation}

We remark that the differential equation for $\A$ is non linear, in the sense that the metric $G$  depends on $\A$
This means that we need to find an initial approximated solution $\A_0$. A natural choice is the solution
of the vector  Laplace equation
$$  \Delta \A_0  = \nabla\phi.$$
Of course this is only an approximated solution $\A_0$, but we can recover a better one $\A_1$ as a solution of
$$d_{\A_0}^*d    \A_1  =  \nabla_{\A_0}\phi,$$
using the subriemmannian operator associated to $\A_0$. Recall that $\nabla_{\A_0} = G\nabla$. From here we start an iteration:
$$d_{\A_1}^*d   \A_2  = \nabla _{\A_1}\phi, \cdots \quad  d_{\A_{j-1}} ^*d  \A_j  = \nabla _{\A_{j-1}}\phi$$
At each step we get a better approximation of the solution, moreover the sequence has a limit $\A= \lim_{j\rightarrow +\infty} \A_j$.
Passing to the limit in the previous expression, we will get:
$$ d_{\A}^*d\A = \nabla_{\A}\phi,$$
so that the limit provides a solution of the nonlinear equation.

\subsubsection{Invariance properties and  choice of the gauge}

The functional $\mathcal {L}$ is invariant with respect to the transformations:
$$h \to h' = h + f,\;\; \phi\to \phi' = \phi + f, \;\; \A\to \A' = \A + d f$$
Indeed  $$d\phi'-dh' =  d\phi-dh,\quad  d\phi'-A' = d\phi-A, \quad d\A'= d\A + ddf= d\A$$
since  $d^2 f= curl df =0$. 
This implies that the functional assumes the same values on $(h,\phi, \A)$ and  $(h',\phi', \A')$
$$\mathcal{L}(h,\phi, \A)=\mathcal{L} (h',\phi', \A').$$

Since the equation is invariant with respect to the choice of the gauge $f$, we can freely choose it, and the choice of the gauge will not affect the value of the Lagrangian. Hence we will make the choice which decouples and symplifies  the system, imposing $a=0$. Since  $\A= \A'-df,$  the expression of $a$ reduces to 
$$a=  g^{12} \partial_{x}( A'_y  - \partial_y f)+  g^{22}\partial_{y} ( A'_y - \partial_y f)   + g^{11} \partial_{x }( A'_x - \partial_x f)   + g^{12} \partial_{y }( A'_x  - \partial_x f),$$
where $f$ is an arbitrary choosen gauge function. 
To obtain $a=0$ we choose the function  $f$ as a solution of $$  g^{22}\partial_{yy} f  + g^{11} \partial_{xx }f + 2 g^{12} \partial_{yx}f=   g^{12} \partial_{x} A'_y  +  g^{22}\partial_{y}  A'_y   + g^{11} \partial_{x } A'_x   + g^{12} \partial_{y } A'_x,$$
that   is  a second order subriemannian differential equation. 

With this choice of the gauge, the second order term of the system reduces to the simpler form: 

$$ d_\A^*d\A =  \left\{ \begin{matrix} \Delta_{\A} A_x      + T_x(\A) \\  \\
 \Delta_{\A} A_y   + T_y(\A)   \end{matrix}\right.$$
where 
$$ 
 \Delta_{\A} f   =  div(  \nabla_\A f)  =  \frac{A_y^2\partial_{xx} f -2 A_xA_y \partial_{xy} f +  A_x^2 \partial_{yy} f }{A_x^2 + A_y^2 }$$ $$
T_x(\vec{A})=  -\partial_{y} g^{11}\partial_{x} A_x - \partial_{y}g^{12} \partial_{y} A_x - \partial_{x}g^{21}\partial_{x}A_y-\partial_{x}g^{22}\partial_{y}A_y 
$$$$
T_y(\vec{A})=- \partial_{y}g^{12} \partial_{x} A_y -  \partial_{y}g^{22}\partial_{y} A_y  - \partial_{x}g^{11} \partial_{x }A_x - \partial_{x}g^{12} \partial_{y }A_x .
$$
In conclusion we can rewrite the Euler Lagrange equation (\ref{Euler}) in the  form

\begin{equation} \label{Euler1}\Bigg\{\begin{matrix}  \Delta \phi =\frac{1}{2}( \Delta h + div (\A)) \\ 
 \Delta_{\A} \A  + T(\A)=- \nabla_\A \phi + \A. \end{matrix} \end{equation}

\subsection{Solution of Euler Lagrange equations}

The implementation of the algorithm consists in solving  the  system of coupled differential equation sequentially.
We first apply the retinex equation to the initial image:
\begin{equation}\label{deltafi}  \Delta \phi = \frac{1}{2} \Delta h\end{equation}
and solve it by convolution
$$ \phi= \frac{1}{2}\Big(\Gamma(x,y)* \Delta h\Big)$$
with the fundamental solution of the 2D Laplacian:
$$\Gamma (x,y) = log(|(x,y)|) .$$

\begin{figure}[H] \label{grad_ind}
\begin{center}
	\includegraphics[width=6.5cm]{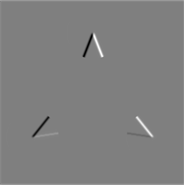}\; \;
	\includegraphics[width=6.5cm]{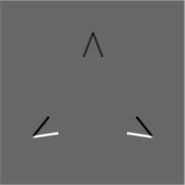}
\end{center}
	\caption{The $x$ and $y$ components of $\nabla h$ related to the Kanizsa triangle inducers. }
	\end{figure}

\begin{figure}[H] \label{A_lin}
	\centering
	\includegraphics[width=5.5cm]{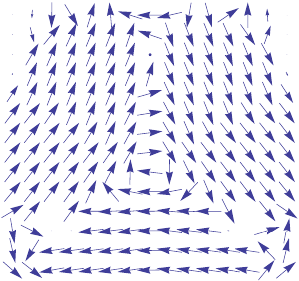}
	\caption{The normalized field $\A_0$ generated by the Kanizsa triangle inducers. }
	\end{figure}

Then we solve the field equation for boundaries propagation. In this first phase we choose $\A=0$ in the right hand side, and the
nonlinear equation reduces to
$$  \Delta _{\A}\A +\vec{T}(\A)= \nabla\phi.$$
As we explained in the previous section, this equation will be solved by linearization, stopped after the first two steps:
\begin{equation}
\left\{
\begin{matrix}\label{system}
  \Delta \A_0  = \nabla\phi\\
  \Delta_{\A_0}\A_1 +\vec{T}(\A) = \nabla\phi.\end{matrix}
\right.
\end{equation}
The solution of the first equation  in \eqref{system} can be computed by convolution
$$\A_0 =\vec{ \Gamma} * \nabla \phi,$$
where $\vec{ \Gamma}(x,y)=(log(|(x,y)|),log(|(x,y)|))$ is the fundamental solution of  the vector Laplacian. When applied to the Kanizsa inducers, the solution $\A_0$ is visualized in Figure \ref{A_lin}, where the triangle inducers have been manually selected (Figure \ref{grad_ind}).

\begin{figure}[H] \label{A_ind}
\begin{center}
	\includegraphics[width=6.5cm]{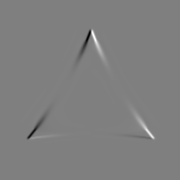}\; \;
	\includegraphics[width=6.5cm]{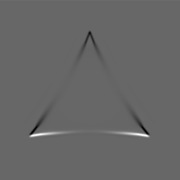}
\end{center}
	\caption{The x and y components of the gauge field $\A_1$ related to the Kanizsa triangle  inducers.  $\A_1$ is an approximation of the field $\A$, solution of the gauge field equation.}
	\end{figure}

The second equation in the same system is a liner degenerate equation. 
If we approximate the matrix $G_{\A_0}$ with its riemannian approximation $\G_\epsilon$, 
it becomes elliptic. 
%According to \eqref{deltadirezionale}, if we denote 
%$(\nu_x, \nu_y) = \Big(  \begin{matrix} \frac{A_x}{\sqrt{A_x^2 + A_y^2}} &   \frac{A_y}{\sqrt{A_x^2 + A_y^2}} \end{matrix}\Big) $ 
%the directional laplacian $ \Delta_{\A_0}$ is a second order derivative in the direction of the vector $(\nu_x, \nu_y)$ and 
%can be represented as 
%$$\Delta_{\A_0} =  
% \frac{A_x^2\partial_{xx}   + 2 A_xA_y \partial_{xy}   +  A_y^2 \partial_{yy}   }{A_x^2 + A_y^2} =
%(\nu_x, \nu_y)
% \Bigg(  \begin{matrix} \partial_{xx} & \partial_{xy}\\  \partial_{xy} & \partial_{yy}  \end{matrix}\Bigg) 
% \Bigg(  \begin{matrix} \nu_x \\ \nu_y \end{matrix}\Bigg) 
%$$
%Hence this operator can be regularized by adding as a small perturbation a second order derivative in the direction  $(-\nu_y, \nu_x)$
%orthogonal to $(\nu_x, \nu_y)$: 
%$$\Delta_{ \epsilon  \A_0 }=  
%(\nu_x, \nu_y)
% \Bigg(  \begin{matrix} \partial_{xx} & \partial_{xy}\\  \partial_{xy} & \partial_{yy}  \end{matrix}\Bigg) 
% \Bigg(  \begin{matrix} \nu_x \\ \nu_y \end{matrix}\Bigg) + \epsilon^2 (-\nu_y, \nu_x)
% \Bigg(  \begin{matrix} \partial_{xx} & \partial_{xy}\\  \partial_{xy} & \partial_{yy}  \end{matrix}\Bigg) 
% \Bigg(  \begin{matrix}- \nu_y \\ \nu_x \end{matrix}\Bigg)
%$$
%Clearly, as  $\epsilon$ tends to $0$ the operator $ \Delta_{ \epsilon  \A_0 }$ provides an approximation of the operator $\Delta_{\A_0} $. 
The solution of the linear elliptic differential equation 
$$\Delta_{ \epsilon  \A_0 } \A_1+\vec{T}(\A)= \nabla \phi$$
provides a good approximation of solution of the second equation in \eqref{system} and it can be obtained by  finite differences methods (centered differences in space). The solution $\A_1$ is shown in Figure \ref{A_ind}.

%The first equation in \ref{system} propagates $\nabla\phi$ isotropically and generates a first appoximated vector field $\A_0$. In Figure \ref{A_lin} the vector field $\A_0$ related to the  Kanizsa inducers is visualized.
%The second equation generates a better approximation $\A_1$ by propagating  $\nabla\phi$  in the direction $\A_0.$  In fig. \ref{A_ind} the components of the vector field $\A_1$ related to the  same inducers is shown. Inducers have been manually selected.

\begin{figure}[H] \label{reconst_fig}
	\centering
         \includegraphics[width=6.5cm]{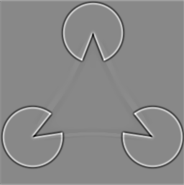}\; \;
	\includegraphics[width=6.5cm]{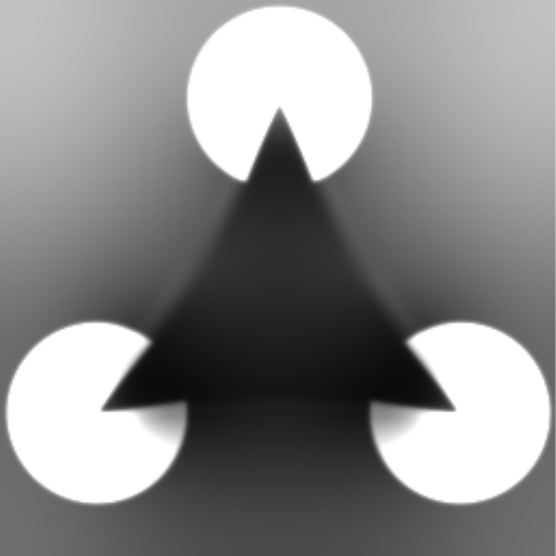}
	\caption{Left: The forcing term $\frac{1}{2}(\Delta h+div(\A_1))$ of the particle equation. Right: The reconstructed Kanizsa triangle as the solution $\phi$ of the particle equation.}
	\end{figure}

Since particle and field equations are coupled we can now solve the complete particle equation
%\begin{equation}\label{feedback}\Delta \phi = \Delta g+\frac{1}{2}(div(\A) )\end{equation}

\begin{equation}\label{complete} 
 \phi= \frac{1}{2}\Gamma(x,y)* \Big(\Delta h+\frac{1}{2}(div(\A_1)\Big)
\end{equation}

again by convolution with the fundamental solution $\Gamma$.
This is a version of the retinex equation able to reconstruct the original image together with the subjective surface. In Figure \ref{reconst_fig} left  it is visualized the forcing term $\frac{1}{2}(\Delta h+div(\A_1))$ of the particle equation while in Figure \ref{reconst_fig} right the solution $\phi$ is shown.

\subsection{A neural implementation in LGN and Cortex}

In order to further support our model, we will discuss how the different terms of the Lagrangian can be implemented in neurophysiological structures.

We recall that the visual signal is first elaborated by the retina whose receptive profiles are  well modeled by the classical  Laplacian of Gaussian:
$$\Delta {\it G}_\sigma = \Delta e^\frac{- x^2}{2 \sigma^2},$$
where $\Delta$ is the standard Laplacian.
We can observe that the same receptive profiles are found in the Lateral Geniculate Nucleus (LGN), that is a copy of the retina but stricly in contact with the visual cortex. 
The action of these receptive fields  on the visual signal can be represented as the output of the neural cell
$$\Delta {\it G}_\sigma * \log I =\Delta h$$
where $g$ is a smoothed version of the  $\log I$ and the logarithmic function is due to the non linearity of the cell response. 
The output of LGN cells is propagated via the horizontal connectivity in LGN. Since this connectivity is isotropic, it can be
modeled with the fundamental solution $\Gamma (x,y)$ of the
2D Laplacian operator.
LGN horizontal connectivity with strength $\Gamma (x,y) $ acts linearly on the feedforward input $h$, giving a total contribution
$$ \phi= \frac{1}{2}\Big(\Gamma* \Delta h\Big).$$
This is exactly the solution
 of the Laplacian
 equation (\ref{deltafi}) of the particle term, implementing the reconstruction of the image from the boundaries. Note that the action of receptive profiles $\Delta h(x,y)$ and the one of LGN horizontal connectivity $\Gamma (x,y) *$ is dual in a differential sense.

The gauge field equation in $\A$ performs boundaries propagation and we will conjecture now how it is implemented at the cortical level.  Simple cells performs stimulus differentiation $\nabla \phi$ that is propagated by horizontal connectivity in the direction of the stimulus orientation\cite{Bosking,Angelucci}. For this reason horizontal connectivity can be modelled by the fundamental solution of  the vector Laplacian $\vec\Gamma(x,y)$ and the total connectivity excited by the stimulus can be accounted as
$$\A_0 =\vec{ \Gamma} * \nabla \phi.$$
Now, the feedforward output of simple cells $\nabla \phi$  is propagated by the excited connectivity $\A_0$ generating the distribution $\A_1$, solution of:
 $$\Delta_{\A_0}\A_1  +\vec{T}(\A)= \nabla\phi.$$
 We have shown in \cite{CittiSarti} that  $\A_1$ is the field tangent to the perceptual association fields measured by Fields, Hayes and Hess in \cite{FHH} and it is cortically  implemented by means of horizontal connectivity propagation.

Finally, the forcing term $\A$ in the particle equation can be interpreted as the feedback of the cortical processing to LGN, showing the strength of the gauge field theory in coupling the activity
of different physiological layers.
Equation (\ref{complete})  is again the retinex equation, but with the feedback from V1 that takes into account illusory boundaries.

%\begin{figure}[H]
%	\centering
%	\includegraphics[width=7.5cm]{fig/kanitza5initial.png}\;\;
%	\includegraphics[width=7.5cm]{fig/kanitza5.png}
%	\caption{the Kaniza pentagone (left) and its reconstruction via this algorithm (right)}
%	\end{figure}

\section{Conclusions}
In this paper we made the effort to  construct a formal field theory of low level vision.  Contemporary instruments of field theory based on Gauge invariances have been used to introduce a complete Lagrangian with its particle, interaction and field terms.
The Lagrangian couples two well known models for lightness and boundary propagation, i.e. the retinex and the neurogeometrical models.  Particularly the problem of modal completion of illusory figures is faced and it is shown how the Euler-Lagrange field equations well represent the process of constitution of the Kanizsa triangle with curved boundaries.
But the interest of the model overcome the formal analogy with particle-fields physical  theory.
In facts it has to be considered as a plausible model for the interaction between different structures of the visual systems, particularly regarding the coupling between the activity of LGN and the one of the visual cortex.
The Gauge Lagrangian formulation seems to be strongly enough to describe both the feedforward and the  feedback processes of low level vision, by keeping the desired invariances.

\section{Appendix} 
We rapidly recall here the definition of differential in the riemannian setting,
 in the special case where det $G$ is a constant, which is the case of the metric in \ref{}.  
We will call $G=g^{-1}$ since $G$ plays the role of inverse of the metric. Then the riemannian scalar product is defined as:

$$<v, w>_g = <g v, w>.$$
If $a$ is a function then we will denote $da$ the usual differential, whose components are $\nabla a = ( \partial_xa ,\partial_ya)$ 
$$ da = \partial_xa dx+\partial_ya dy.$$
The gradient of the function $a$ in the metric $g$ is defined as $$\quad  \nabla_g  a= G\nabla a =  \left(\begin{matrix} g^{11} & g^{12}  \\ g^{21} & g^{22}\end{matrix} \right) \left(\begin{matrix} \partial_xa\\ \partial_ya\end{matrix} \right)$$
In the sequel we will denote $((\nabla_g a) _x, (\nabla_g a) _y)$ its components. 
The laplacian  is expressed as 
$$\Delta_g a = div(\nabla_g a).$$

If
$\A =A_x dx + A_y dy$, then 
$$d\A = curl (\A) = (\partial_xA_y- \partial_yA_x) \;\;dx\wedge dy$$
and the Laplacian is not the laplacian of the two components in general, but can be expressed in terms of the $d$ and $d^*$ operators, which we will now define. Since $d\A $ is a 2-form,  first recall that 
for a general 2-form  $\beta =b dx\wedge dy$

$$d^*\beta =- (\nabla_g b)_y  dx + (\nabla_g b)_x  dy  $$
$$=   - (g^{21}\partial_xb  + g^{22}\partial_yb) dx +  (  g^{11} \partial_xb   + g^{12} \partial_yb )dy $$
(formally its components are $ \nabla_g ^\bot b$, so that 
%It is well known that the Euler Lagrangian equation of $$\int |d\A|^2 dxdy $$
%is $$d^*d\alpha=0$$ which, from the previous computations, is
%
%\vfill\eject
so that, while applying to $d\alpha$, 

\begin{equation} \label{distar}d^*d\A=   \nabla_g ^\bot  curl (\alpha) =- (\nabla_g (\partial_xA_y- \partial_yA_x))_y  dx + (\nabla_g (\partial_xA_y- \partial_yA_x))_x  dy.\end{equation}

The vanishig condition of this expression (which will be used in section ??)  is a system in two variables. 

$$  \left\{ \begin{matrix}( \nabla_g \partial_yA_x)_y  - ( \nabla_g \partial_xA_y)_y     =0\\  \\ 
 (\nabla_g \partial_xA_y)_x - (\nabla_g \partial_yA_x)_x   =0 \end{matrix}\right.$$

We can now exchange the order of differentiation: 
we call $ [ \nabla_g,  \partial_x]=    \nabla_g \partial_x - \partial_x  \nabla_g$

$$  \left\{ \begin{matrix} \partial_y (\nabla_g A_x)_y  +[ \partial_y, (\nabla_g)_y] A_x   - \partial_x ( \nabla_g A_y)_y   
- [\partial_x ,  (\nabla_g)_y]  A_y  =0\\  \\ 
 \partial_x(\nabla_g A_y)_x  +[ \partial_x, (\nabla_g)_x] A_y   - \partial_y(\nabla_g A_x)_x   -  [\partial_y, (\nabla_g )_x]A_x  =0 \end{matrix}\right.$$

Now we note that  $div((\nabla_g A_x) =  \partial_x (\nabla_g A_x)_x + \partial_y (\nabla_g A_x)_y$

$$  \left\{ \begin{matrix} div(  \nabla_g A_x)     +[ \partial_y, (\nabla_g)_y] A_x   
- [\partial_x ,  (\nabla_g)_y] A_y  - \partial_x(  ( \nabla_g  A_x)_x    +  ( \nabla_g A_y)_y )=0\\  \\ 
div(  \nabla_g A_y)    
  +[ \partial_x, (\nabla_g)_x] A_y    -  [\partial_y, (\nabla_g )_x]A_x  - \partial_y((\nabla_g A_x)_x  + \nabla_g A_y)_ y)=0 \end{matrix}\right.$$

Now we give a name of each term in this expression,   
\begin{equation}\label{T12h}
\left\{\begin{matrix}div(  \nabla_g  A_i) =  \Delta_g  A_i\quad \quad \quad \quad\quad \quad \quad \quad \quad\quad \quad \quad \quad\quad\quad \quad \quad \quad\quad \quad \quad  \quad\quad \\
T_x(\vec{A})=[ \partial_y, (\nabla_g)_y] A_x   
- [\partial_x ,  (\nabla_g)_y] A_y =\quad \quad \quad \quad\quad \quad \quad \quad \quad\quad \quad \quad \quad\\ \quad \quad \quad \quad\quad \quad \quad \quad = -\partial_{y} g^{11}\partial_{x} A_x - \partial_{y}g^{12} \partial_{y} A_x - \partial_{x}g^{21}\partial_{x}A_y-\partial_{x}g^{22}\partial_{y}A_y 
\\
T_y(\vec{A})= [ \partial_x, (\nabla_g)_x] A_y-  [\partial_y, (\nabla_g )_x]A_x =\quad \quad \quad \quad\quad \quad \quad \quad \quad\quad \quad \quad \quad\\ \quad \quad \quad \quad\quad \quad \quad \quad =- \partial_{y}g^{12} \partial_{x} A_y -  \partial_{y}g^{22}\partial_{y} A_y  - \partial_{x}g^{11} \partial_{x }A_x    - \partial_{x}g^{12} \partial_{y }A_x 
\\
a=  (\nabla_g A_x)_x  + (\nabla_g A_y)_ y= g^{12} \partial_{x} A_y +  g^{22}\partial_{y} A_y   + g^{11} \partial_{x }A_x   + g^{12} \partial_{y }A_x 
\end{matrix} \right.\end{equation}

Finally we conclude that the first variation of the functional
$$\int |d\A|_\A^2$$

 can be expressed as 

$$  d^*d\A =  \nabla_g^\bot curl (\A) = \left\{ \begin{matrix} \Delta_g A_x     - \partial_x a + T_x(\A) \\  \\
 \Delta_g A_y     - \partial_y a + T_y(\A)   \end{matrix}\right.$$

\end{document}